\documentclass[journal]{IEEEtran}
\usepackage{amsmath,amssymb}
\usepackage[dvips]{graphicx}
\usepackage{amsfonts}
\usepackage[mathscr]{eucal}
\usepackage{mathtools, nccmath}
\usepackage{latexsym}
\usepackage{amsthm}
\usepackage{exscale}
\usepackage[mathscr]{eucal}
\usepackage{bm}
\usepackage[dvipsnames]{color}
\usepackage{cases}
\usepackage{color}
\usepackage{epsfig}
\usepackage[ruled]{algorithm2e}
\usepackage[verbose,nospace,sort]{cite}
\usepackage{tabularx,ragged2e}
\usepackage{multirow}
\usepackage{multicol}
\usepackage{balance}
\usepackage{caption}
\usepackage{subcaption}
\usepackage{setspace}
\usepackage{multirow}
\usepackage{multicol,afterpage,wrapfig}
\usepackage{pifont}
\usepackage{array}

\usepackage{algpseudocode}  
\usepackage{amsmath}

\graphicspath{{./figs/}}

\newtheorem{lemma}{Lemma}

\begin{document}

\title{Intelligent Reflecting Surface Assisted Secrecy Communication: Is Artificial Noise Helpful or Not?}

\author{{Xinrong Guan, \IEEEmembership{Member, IEEE,}
		Qingqing Wu, \IEEEmembership{Member, IEEE,}
		and Rui Zhang, \IEEEmembership{Fellow, IEEE}
		\thanks{X. Guan is with the College of Communications Engineering, Army Engineering University of PLA, Nanjing, 210007, China. Q. Wu and R. Zhang are with the Department of Electrical and Computer Engineering, National University of Singapore, 117583, Singapore (e-mails: geniusg2017@gmail.com, {elewuqq, elezhang}@nus.edu.sg). }}
} 

\maketitle

\begin{abstract}
In this letter, we investigate whether the use of artificial noise (AN) is helpful to enhance the secrecy rate of an intelligent reflecting surface (IRS) assisted wireless communication system. Specifically, an IRS is deployed nearby a single-antenna receiver to assist in the transmission from a multi-antenna transmitter, in the presence of multiple single-antenna eavesdroppers. Aiming to maximize the achievable secrecy rate, a design problem for jointly optimizing  transmit beamforming with AN or jamming and IRS reflect beamforming is formulated, which is however difficult to solve due to its non-convexity and coupled variables. We thus propose an efficient algorithm based on alternating optimization to solve the problem sub-optimally. Simulation results show that incorporating AN in transmit beamforming is beneficial under the new setup with IRS reflect beamforming. In particular, it is unveiled that the IRS-aided design without AN even performs worse than the AN-aided design without IRS as the number of eavesdroppers near the IRS increases.
\end{abstract}
%

%
\IEEEpeerreviewmaketitle
\vspace{-2mm}
\section{Introduction}
Recently, intelligent reflecting surface (IRS) has been proposed as a key enabling technology for achieving a smart and reconfigurable signal propagation environment in future wireless networks \cite{QQ}, \cite{Basar}. Specifically, IRS is a metasurface composed of a large number of low-cost passive reflecting elements. By adaptively adjusting the reflection amplitude and/or phase shift of each element at an IRS, the strength and direction of the electromagnetic wave becomes highly controllable, whereby the reflected signal can be intentionally enhanced or weakened at different receivers. Moreover, IRS consumes much less power than traditional active transceivers/relays since it merely reflects signals without injecting any power for amplification \cite{QQ_1}. As a new promising solution to achieve high beamforming gain with very low hardware/energy cost, IRS has been applied in various wireless applications such as coverage extension, interference cancellation, energy efficiency enhancement, and so on (see \cite{QQ} and the references therein). 

On the other hand, physical layer security has been thoroughly investigated as a complement to higher-layer encryption techniques, for ensuring wireless security from an information-theoretic perspective. By exploiting the spatial degrees of freedom (DoF), transmit beamforming can be designed to direct the signal towards the legitimate user and meanwhile degrade the reception at the eavesdropper, so that the secrecy rate is maximized. An effective approach to enhance the secrecy beamforming is via combining jamming or artificial noise (AN) with it, which is particularly helpful when the number of eavesdroppers is larger than that of transmit antennas \cite{LQ_2013}. This is because the transmitter in this case lacks sufficient DoF to send the legitimate signal into the null space of all the eavesdroppers' channels, thus rendering the standalone transmit beamforming ineffective and the use of AN necessary. 

Thanks to its capability of configuring wireless channels smartly, IRS has great potential in enhancing physical layer security and IRS-assisted secrecy communication was recently investigated in \cite{Guangchi,Shen,Yu,LYC}. Via jointly designing the active transmit beamforming and the passive reflect beamforming of the IRS that is usually deployed near the legitimate receiver, the achievable secrecy rate can be significantly improved. However, the above works mainly focused on the joint beamforming design using various different optimization methods, while the transmit jamming with AN was not considered therein. To the authors' best knowledge, it still remains an open problem whether AN is helpful under the new setup with an IRS deployed to assist in the secure communication.

This thus motivates the current work to investigate the joint transmit beamforming with AN and IRS reflect beamforming in an IRS-assisted secrecy communication system, as shown in Fig. 1. We aim to maximize the achievable secrecy rate of the considered system and thereby investigate: (1) whether the additional DoF brought by the IRS can have any impact on the necessity of using AN in the joint beamforming design; and (2) under what conditions AN is most helpful. Simulation results show that even with the help of IRS reflect beamforming, incorporating jamming or AN is still effective to improve the secrecy rate, especially when the transmit power is large for achieving high secrecy rate and/or the number of eavesdroppers increases. It is also unveiled that as the number of reflecting elements increases, the performance gain brought by AN is roughly constant when the eavesdroppers are far away from the IRS, but decreases when the eavesdroppers are located near the IRS.


\vspace{-1mm}
\section{System Model and Problem Formulation}
\vspace{-1mm}
\subsection{System Model}\vspace{-1mm}
As shown in Fig. 1, we consider a wireless communication system where a legitimate transmitter (Alice) intends to send confidential information to a legitimate receiver (Bob) with the help of an IRS (Rose) that is deployed nearby Bob, against $K$ eavesdroppers{\footnote {Eves are assumed to be other users in this network, but they are not intended to receive this confidential information.}} (Eves) that are arbitrarily distributed in the system. Suppose that Bob and all Eves are equipped with a single antenna, while the number of antennas at Alice and that of reflecting elements at Rose are denoted by $M$ and $N$, respectively. The baseband equivalent channels from Alice to Rose, Bob and Eve $k$ (the $k$-th eavesdropper) are denoted by ${{\mathbf{H}}_{ar}} \in {{\mathbb{C}}^{N \times M}}$, ${{\bf{h}}{_{ab}^H}} \in {{\mathbb{C}}^{1 \times M}}$ and ${{\mathbf{h}}^H_{ae_k}} \in {{\mathbb{C}}^{1 \times M}}$, respectively, while those from Rose to Bob and Eve $k$ are denoted by  ${{\mathbf{h}}^H_{rb}} \in{\mathbb{C}^{1 \times N}}$ and ${{\mathbf{h}}^H_{re_k}} \in{\mathbb{C}^{1 \times N}}$, respectively. Let ${\mathbf{\Phi }} = \text{diag}\left( {{e^{j{\theta _1}}},{e^{j{\theta _2}}},....,{e^{j{\theta _N}}}} \right)$ represent the diagonal phase-shifting matrix of Rose, where in its main diagonal, $\theta_n \in [0,2\pi)$ is the phase shift on the combined incident signal by its $n$-th element, $n=1,...,N$ \cite{QQ_1}. The composite Alice-Rose-Bob/Eve $k$ channel is then modeled as a concatenation of three components, namely, the Alice-Rose link, Rose's reflection with phase shifts, and Rose-Bob/Eve $k$ link. In addition, the quasi-static flat-fading model is assumed for all channels. To characterize the performance limit of the considered IRS-assisted secrecy communication system, we assume that the channel state information (CSI) of all channels involved is perfectly known at Alice and Rose for their joint design of transmit/reflect beamforming and jamming, based on the various channel acquisition methods discussed in \cite{QQ} and \cite{Zheng2019}.

The transmitted signal from Alice is given by \vspace{-1.5mm}
\begin{equation}
{\mathbf{x}} = {{\mathbf{f}}_1}s + {{\mathbf{f}}_2}a,
	\vspace{-0.5mm}
\end{equation}
where $s\sim {\mathcal{CN}} \left( {0,1} \right) $ and $a\sim {\mathcal{CN}}\left( {0,1} \right)$ denote the independent information and jamming/AN signals, respectively, while ${{\mathbf{f}}_1\in {\mathbb{C}^{M \times 1}}}$ and ${{\mathbf{f}}_2\in {\mathbb{C}^{M \times 1}}}$ denote the beamforming and jamming vectors, respectively. Assuming that Alice has a maximum transmit power budget $P_{\rm max}$, we have ${\bf{f}}_1^H{\bf{f}}_1+{\bf{f}}_2^H{\bf{f}}_2 \le P_{\rm max}$. The signal received at Bob or Eve $k$ is then given by\vspace{-0mm}
\begin{equation}
{y_i} \!=\! \left( {{\mathbf{h}}_{ai}^H{\text{ \!+\! }}{\mathbf{h}}_{ri}^H{\mathbf{\Phi H}}_{ar}} \right)\left( {{{\mathbf{f}}_1}s \!+\! {{\mathbf{f}}_2}a} \right) \!+\! {n_i}, ~{i \in \{b,e_k\}},\vspace{-0mm}
\end{equation}
where ${n_i} \sim \mathcal{CN}\left( {0,\sigma_0^2} \right)$ is the complex additive white Gaussian noise (AWGN). Let ${{\mathbf{v}}^H} = \left[ {{v_1},{v_2},...,{v_N}} \right]$ where ${v_n} = {e^{j{\theta _n}}}$, $\forall n$. By changing variables as ${{\mathbf{h}}_{ri}^H{\mathbf{\Phi H}}_{ar} = {{\mathbf{v}}^H}{{\mathbf{H}}_{ari}}}$ where ${{\bf{H}}_{ari}} = \text{diag}\left( {{\bf{h}}_{ri}^H} \right){\bf{H}}_{ar}$, the signal-to-interference-plus-noise ratio (SINR) at Bob or Eve $k$ can be derived as\vspace{-0mm}
\begin{equation}
	\small
	\vspace{-0mm}
	\label{SINR}
	{\gamma _i} = \frac{{{\gamma _0}{{\left| {{{{\mathbf{\tilde v}}}^H}{{\mathbf{H}}_i}{{\mathbf{f}}_1}} \right|}^2}}}{{{\gamma _0}{{\left| {{{{\mathbf{\tilde v}}}^H}{{\mathbf{H}}_i}{{\mathbf{f}}_2}} \right|}^2} + 1}},~{i \in \{b,e_k\}},
\vspace{-0mm}
\end{equation}
where ${\gamma _0} = {{{1}}}/{{\sigma_0^2}}$,  $\small{{\mathbf{H}}_i} = \left[ {\begin{array}{*{20}{c}}
	{{{\mathbf{H}}_{ari}}} \\ 
	{{\mathbf{h}}_{ai}^H} 
	\end{array}} \right]$, ${{{\mathbf{\tilde v}}}^H} = {e^{j\varpi }}\left[ {{{\mathbf{v}}^H},1} \right]$ and ${\varpi }$ is an arbitrary phase rotation.
\vspace{-1mm}
\subsection{Problem Formulation}
\vspace{0.5mm}
We aim to maximize the achievable secrecy rate via a joint design of the transmit beamforming and jamming at Alice and the reflect beamforming at Rose, subject to the total power constraint at Alice. As such, the optimization problem is formulated as
\vspace{-1mm}
\begin{equation*}	\vspace{-1.5mm}
\label{P_0}
	\begin{split}
	\left( {{\text{P0}}} \right) :
	\mathop {\max }\limits_{{\bf{f}}_1,{\bf{f}}_2,{\bf{v}}} ~~~
	& \left\{R_b-\mathop {\max }\limits_k ~R_{e_k}\right\}\\	
	{{\rm  s.t.}}~~~~&{{\bf{f}}_1^H}{\bf{f}}_1 +{{\bf{f}}_2^H}{\bf{f}}_2\le {P_{\max }},\\
	&\left| {{v_n}} \right| = 1, n=1,...,N,
	\end{split}
\end{equation*}
where $R_b=\log \left( {{1 + {\gamma _b}}} \right)$ and $R_{e_k}=\log \left( {{1 + {\gamma _{e_k}}}} \right)$ are the achievable rates in bits/second/Hertz (bps/Hz) for Bob and Eve $k$, respectively, and $\log(x)$ denotes the base-2 logarithm of $x$. (P0) is difficult to solve due to the non-concave objective function as well as the coupled optimization variables. However, we observe that the resultant problems can be efficiently solved when one of $({\bf{f}}_1,{\bf{f}}_2)$ and ${\bf{v}}$ is fixed. This thus motivates us to propose an alternating optimization based algorithm to solve (P0) sub-optimally, by iteratively optimizing one of $({\bf{f}}_1,{\bf{f}}_2)$ and ${\bf{v}}$ with the other being fixed at each iteration until convergence is reached, as detailed in the next section. 
\begin{figure}[t]
	\label{systemmodel}
	\centering{\includegraphics[width=3.3in]{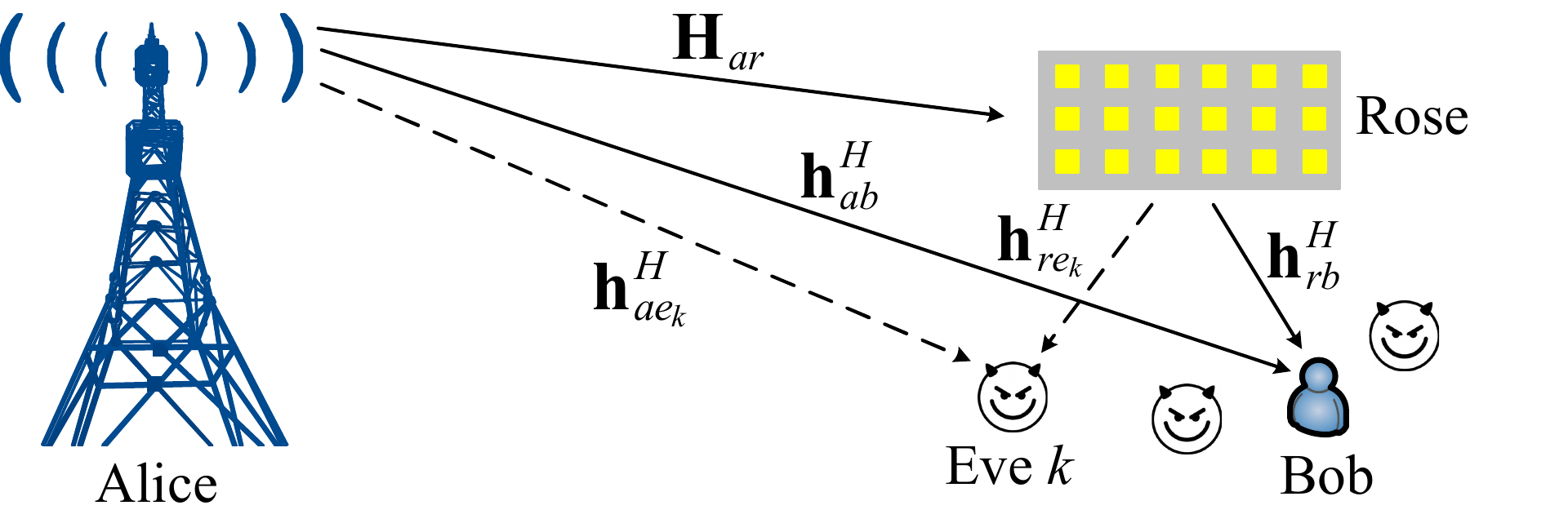}}
	\caption{IRS-assisted wireless secrecy communication.}
	\vspace{-1mm}
\end{figure}
\vspace{-1mm}
\section{Joint Design of Beamforming and Jamming}
\subsection{Optimizing ${\bf{f}}_1$ and ${\bf{f}}_2$ for Given ${\bf{v}}$}
\vspace{-0mm}
For given ${\bf{v}}$, we denote ${{{\mathbf{\tilde H}}}_b} = {{{\mathbf{\tilde h}}}_b}{\mathbf{\tilde h}}_b^H$ and ${{{\mathbf{\tilde H}}}_{e_k}} = {{{\mathbf{\tilde h}}}_{e_k}}{\mathbf{\tilde h}}_{e_k}^H$, where ${\mathbf{\tilde h}}_b^H = {{{\mathbf{\tilde v}}}^H}{{\mathbf{H}}_b}$ and ${\mathbf{\tilde h}}_{e_k}^H = {{{\mathbf{\tilde v}}}^H}{{\mathbf{H}}_{e_k}}$ can be viewed as the
effective channels from Alice to Bob and Eve $k$, respectively, by combining the direct channel and the IRS-reflected channel. Then, (P0) can be transformed to the following problem
\vspace{-0.5mm}
\begin{equation*}\vspace{-0mm}
\label{P_1.1}
\small
\begin{split}
\left( {{\text{P1.1}}} \!\right) \!:
\mathop {\max }\limits_{{\bf{f}}_1,{\bf{f}}_2} \
& \log\!\!\left(\!\!1\!\!+\!\!\frac{{{\gamma _0}{{\left| {{\mathbf{\tilde h}}_b^H{{\mathbf{f}}_1}} \right|}^2}}}{{{\gamma _0}{{\!\left| {{\mathbf{\tilde h}}_b^H{{\mathbf{f}}_2}} \right|\!}^2}\! \!+\!\! 1}}\!\!\right)\!\!\!-\!\mathop {\max }\limits_k\ \log\!\!\left(\!\!1\!\!+\!\!\frac{{{\gamma _0}{{\left| {{\mathbf{\tilde h}}_{e_k}^H{{\mathbf{f}}_1}} \right|}^2}}}{{{\gamma _0}{{\!\left| {{\mathbf{\tilde h}}_{e_k}^H{{\mathbf{f}}_2}} \right|\!}^2} \!\!+\!\! 1}}\!\!\right)\!\!\\	
{\rm  s.t.}~~&{{\bf{f}}_1^H}{\bf{f}}_1 +{{\bf{f}}_2^H}{\bf{f}}_2\le {P_{\max }}.
\end{split}
\end{equation*}\vspace{-1.5mm}\\
Note that ${| {{\mathbf{\tilde h}}_i^H{{\mathbf{f}}_1}} |}^2={\mathbf{Tr}}{({{{\mathbf{\tilde H}}}_i} {{{\mathbf{f}}_1}{{\mathbf{f}}_1^H} } )}$ and ${| {{\mathbf{\tilde h}}_i^H{{\mathbf{f}}_2}} |}^2={\mathbf{Tr}}{({{{\mathbf{\tilde H}}}_i} {{{\mathbf{f}}_2}{{\mathbf{f}}_2^H} } )}$, $i \in \{b, {e_k}\}$. Define two matrices as ${{\mathbf{F}}_1}={{\mathbf{f}}_1}{{\mathbf{f}}_1^H}$ and ${{\mathbf{F}}_2}={{\mathbf{f}}_2}{{\mathbf{f}}_2^H}$. Then it follows that ${{\mathbf{F}}_1} \succeq 0$, ${{\mathbf{F}}_2} \succeq 0$ and ${\rm rank}({{\mathbf{F}}_1})={\rm rank}({{\mathbf{F}}_2})=1$. Since the rank-1 constraints are non-convex, we apply the semidefinite relaxation (SDR) to relax these constraints. As a result, (P1.1) is reduced to\vspace{-0.5mm}
\begin{equation*}
\vspace{-1.5mm}
 \label{P_1.2}
 \small
 \begin{split}
	 \left( {{\text{P1.2}}} \!\right) \!:
	 \mathop {\max }\limits_{{\bf{f}}_1,{\bf{f}}_2} \
	 & \!\log\!\!\!\left(\!\!1\!\!+\!\!\frac{{{\gamma _0}{{\mathbf{Tr}}{({{{\mathbf{\tilde H}}}_b} {{{\mathbf{F}}_1} } \!)\!}}}}{{{\gamma _0}{\!{\mathbf{Tr}}\!{({{{\mathbf{\tilde H}}}_b} {{{\mathbf{F}}_2} } \!)\!}}\! +\! 1}}\!\!\right)\!\!\!-\!\mathop {\max }\limits_k\ \!\log\!\!\!\left(\!\!1\!\!+\!\!\frac{{{\gamma _0}{{\mathbf{Tr}}{({{{\mathbf{\tilde H}}}_{e_k}} {{\!{\mathbf{F}}\!_1} } \!)\!}}}}{{{\gamma _0}{\!{\mathbf{Tr}}\!{({{{\mathbf{\tilde H}}}_{e_k}} {{\!{\mathbf{F}}\!_2} } \!)\!}}\! +\! 1}}\!\!\right)\!\!\\	
	{\rm  s.t.}\,~&\left( {{{\mathbf{F}}_1},{{\mathbf{F}}_2}} \right) \in \mathcal{F},
 \end{split}\vspace{-1.5mm}
\end{equation*}
where	\vspace{-1.5mm}
\begin{equation*}
\mathcal{F} \! =\! \left\{ \left( {{{\mathbf{F}}_1},{{\mathbf{F}}_2}} \right)\left| {{\text{Tr}}\left( {{{\mathbf{F}}_1} \!+\! {{\mathbf{F}}_2}} \right) \!\le\! {P_{\max }},} \right.{{\mathbf{F}}_1} \succeq 0,{{\mathbf{F}}_2} \succeq 0 \right\}	\vspace{-1mm}
\end{equation*}
is the feasible set for $({{\bf{F}}_1}, {{\bf{F}}_2)}$. 
However, (P1.2) is still difficult to solve since the objective function is not jointly concave with respect to (w.r.t.) ${\bf{F}}_1$ and ${\bf{F}}_2$, which are non-trivially coupled too. To overcome these difficulties, we resort to the following lemma \cite{LQ_2013}.
\begin{lemma}\vspace{-0.5mm}
	Consider the function $\varphi \left( t \right) =  - tx + \ln t + 1$ for any $x>0$. Then, we have\vspace{-1.5mm}
	\begin{equation}
	- \ln x=\mathop {\max }\limits_{t > 0} \varphi \left( t \right)   ,\vspace{-0.5mm}
	\end{equation} and the optimal solution is $t=1/x$.\vspace{-0mm}
\end{lemma}
Lemma 1 provides an upper bound for $\varphi \left( t \right)$, and this bound is tight when $t=1/x$. By applying Lemma 1 and setting $x={{\gamma _0}{\mathbf{Tr}}( {{{\mathbf{\tilde H}}}_b} {{{\mathbf{F}}_2} } ) + 1}$ and $t=t_b$, $R_b$ can be written as\vspace{-1mm}
{\small
\begin{align}
 	\!\!{R_b}\ln2 &\!\!=\!{\ln \!\left(\! {{\gamma _0}{\mathbf{Tr}}\!\left(\! {{{{\mathbf{\tilde H}}}_b}\left( {{{\mathbf{F}}_1} \!+\! {{\mathbf{F}}_2}} \right)} \!\right)\! \!+\! 1} \right) \!-\! \ln \!\left(\! {{\gamma _0}{\mathbf{Tr}}\!\left(\! {{{\mathbf{\tilde H}}}_b} {{{\mathbf{F}}_2} }  \right) \!\!+\!\! 1} \right) \!}\! \notag\\
 	&\!=\! \mathop {\max }\limits_{{t_b} > 0}{ {\varphi _b}\left( {{{\mathbf{F}}_1},{{\mathbf{F}}_2},{t_b}} \right)\!},\label{C_b}\vspace{-2mm}    
\end{align}
}
\vspace{-4mm}\\
where \vspace{-2mm}
{\small
\begingroup
\addtolength{\jot}{-1mm}
\begin{align}
    {\varphi _b}\left( {{{\mathbf{F}}_1},{{\mathbf{F}}_2},{t_b}} \right) = &\ln \left( {{\gamma _0}{\mathbf{Tr}}\left( {{{{\mathbf{\tilde H}}}_b}\left( {{{\mathbf{F}}_1} + {{\mathbf{F}}_2}} \right)} \right) + 1} \right) - \notag \\
	&{t_b}\left( {{\gamma _0}{\mathbf{Tr}}\left( {{{{\mathbf{\tilde H}}}_b}{{\mathbf{F}}_2}} \right) + 1} \right) + \ln {t_b} + 1. \vspace{-0mm} 
\end{align}
\endgroup
}\vspace{-3mm}\\
Similarly, by setting $x={{\gamma _0}{\mathbf{Tr}}\left( {{\!{{\mathbf{\tilde H}}_{e_k}}\!}\left( {{{\mathbf{F}}_1} + {{\mathbf{F}}_2}} \right)} \right) + 1}$ and $t={t_{e_k}}$, ${R_{e_k}}$ can be expressed as\vspace{-1mm}
{\small
\begingroup
\addtolength{\jot}{-0.5mm}
\begin{align}
	{R_{e_k}} {\ln2}&\!={\! \ln \!\left(\! {{\gamma _0}\!{\mathbf{Tr}}\!\left( {{\!{{\mathbf{\tilde H}}_{e_k}}\!}\left( {{{\mathbf{F}}_1} \!\!+\!\! {{\mathbf{F}}_2}} \right)} \!\right)\! \!\!+\!\! 1} \!\right) \!\!\!-\!\! \ln \!\left(\! {{\gamma _0}\!{\mathbf{Tr}}\!\left( {\!{{\mathbf{\tilde H}}_{e_k}}\!} {{{\mathbf{F}}_2} }  \!\right)\! \!\!+\!\! 1} \!\right)\!\! }\notag\\
	&\!=\! \mathop {\min }\limits_{{t_{e_k}>0}}{{\varphi _{e_k}}\left({{{\mathbf{F}}_1},{{\mathbf{F}}_2},{t_{e_k}}} \right)},\label{C_ek}	
\end{align}
\endgroup
}
where
\vspace{-2mm}
{\small
\begingroup
\addtolength{\jot}{-0.5mm}
\begin{align}
\vspace{-2mm}
	\!{\varphi _{e_k}}\!\left( {{{\mathbf{F}}_1},{{\mathbf{F}}_2},{t_{e_k}}} \right)\! = &{t_{e_k}}\!\left(\! {{\gamma _0}{\mathbf{Tr}}\left( {{{{\mathbf{\tilde H}}}_{e_k}}\left( {{{\mathbf{F}}_1} + {{\mathbf{F}}_2}} \right)} \right) \!+ \!1} \right) - \! \notag\\
	&\ln \left( {{\gamma _0}{\mathbf{Tr}}\left( {{{{\mathbf{\tilde H}}}_{e_k}}{{\mathbf{F}}_2}} \right) \!+\! 1} \right)\!-\! \ln {t_{e_k}}\!-\! 1.  
\end{align}
\endgroup
}
\vspace{-4mm}\\
Therefore, following Sion's minimax theorem \cite{Sion1958}, (P1.2) can be rewritten as\vspace{-2mm}
\begingroup
\addtolength{\jot}{-1mm}
\begin{equation*}
\small
\vspace{-0.5mm}
\label{P_1.3}
	\begin{split}
	\!\!\left( {{\text{P1.3}}} \right)\!:\!\mathop {\max }\limits_{{\!{\mathbf{F}}\!_1},{{\mathbf{F}}_2},{t_b},{t_{{\!e_k\!}}}} \!&\!\left\{\!{\varphi _b}\!\left(\,  {{\!{\mathbf{F}}\!_1},{{\mathbf{F}}_2},{t_b}} \right)\! \!-\! \mathop {\!\max\! }\limits_k \ {\!\varphi _{{e_k}}\!}\!\left(\,  {{\!{\mathbf{F}}\!_1},{{\mathbf{F}}_2},\!{t_{{e_k}}}\!} \right)\!\right\}\\ 
	{\rm  s.t.}~~~&\left( {{{\mathbf{F}}_1},{{\mathbf{F}}_2}} \right) \in \mathcal{F}, \\
	&{t_b} > 0,{t_{e_k}} > 0, k=1,...,K.  \!
	\end{split}
	\vspace{-1mm}
\end{equation*}
\endgroup
Note that the constant ``$\ln2$'' is omitted in the objective function without loss of optimality. It can be shown that (P1.3) is convex w.r.t. either $\left({{\bf{F}}_1}, {{\bf{F}}_2}\right)$ or $\left(t_b,t_{e_k}\right)$. Thus, it can be solved by applying the alternating optimization technique. 
\setlength{\intextsep}{0pt}
\setlength{\textfloatsep}{4mm} 
\begin{algorithm}[t]
	\DontPrintSemicolon
	\LinesNumbered	
	\SetAlgoSkip{smallskip}
	\SetAlgoHangIndent{0.1em}
	\caption{Alternating optimization for solving (P1.1)}
	\label{Algorithm1}	
	\KwIn{$P_{\rm max}$, $\gamma_0$, ${{\mathbf{\tilde v}}}$, ${\bf{H}}_{b}$, ${\bf{H}}_{{e_k}}$.}
	\KwOut{${{\bf{f}}_1}$, ${{\bf{f}}_2}$.}
	Initialize ${{\bf{f}}_1}$ and ${{\bf{f}}_2}$ according to the maximum transmit power constraint ${\bf{f}}_1^H{\bf{f}}_1+{\bf{f}}_2^H{\bf{f}}_2 \le P_{\rm max}$. \\
	Set $m=1$, ${{\bf{F}}_1^{(0)}}={{\bf{f}}_1}{{\bf{f}}_1^{H}}$, ${{\bf{F}}_2^{(0)}}={{\bf{f}}_2}{{\bf{f}}_2^{H}}$, ${\mathbf{\tilde h}}_b^H = {{{\mathbf{\tilde v}}}^H}{{\mathbf{H}}_b} $, ${\mathbf{\tilde h}}_{e_k}^H = {{{\mathbf{\tilde v}}}^H}{{\mathbf{H}}_{e_k}} $, ${{{\mathbf{\tilde H}}}_b} = {{{\mathbf{\tilde h}}}_b}{\mathbf{\tilde h}}_b^H$, and ${{{\mathbf{\tilde H}}}_{e_k}} = {{{\mathbf{\tilde h}}}_{e_k}}{\mathbf{\tilde h}}_{e_k}^H$.\\
	\Repeat{{\rm the objective value of (P1.1) reaches convergence.}}
	{
		With given $\!{{\bf{F}}_1^{(m-1)}}\!$ and ${{\bf{F}}_2^{(m-1)}}$, find the optimal $t_b^{(m)}$  and $t_{e_k}^{(m)}$ according to \eqref{t_b} and \eqref{t_e}, respectively. \\
		With given $t_b^{(m)}$ and $t_{e_k}^{(m)}$, find the optimal ${{\bf{F}}_1^{(m)}}$ and ${{\bf{F}}_2^{(m)}}$ by solving (P1.5).  \\
		Update $m=m+1$. 	
	}
	Recover ${{\bf{f}}_1}$ and ${{\bf{f}}_2}$ from ${{\bf{F}}_1}$ and ${{\bf{F}}_2}$, respectively.
\end{algorithm}

According to Lemma 1, the optimal $\left(t_b,t_{e_k}\right)$ for fixed $({{\bf{F}}_1}, {{\bf{F}}_2)}$ can be derived in closed-forms as
\vspace{-1mm}
{\small
\begingroup
\addtolength{\jot}{-1mm}
\begin{align}
\vspace{-2mm}
	t_b^*&=\left({{\gamma _0}{\mathbf{Tr}}\!\left(\! {{{\mathbf{\tilde H}}_{b}}} {{{\mathbf{F}}_2} }  \right) \!+\! 1}\right)^{-1},\label{t_b}\\
	t_{e_k}^*&=\left( {{\gamma _0}{\mathbf{Tr}}\!\left(\! {{{{\mathbf{\tilde H}}}_{e_k}}\left( {{{\mathbf{F}}_1} \!+\! {{\mathbf{F}}_2}} \right)} \!\right)\! \!+\! 1} \right)^{-1}.\label{t_e}
\end{align}
\endgroup
}\vspace{-4mm}\\
On the other hand, the optimal $({{\bf{F}}_1}, {{\bf{F}}_2)}$ for given $\left(t_b^*,t_{e_k}^*\right)$ can be obtained by solving 
\vspace{-1mm}
\begingroup
\addtolength{\jot}{-0.5mm}
\begin{equation*}
\small
\vspace{-0.5mm}
\label{P_1.4}
	\begin{split}
	\!\!\left( {{\text{P1.4}}} \right)\!:\mathop {\max }\limits_{{{\mathbf{F}}_1},{{\mathbf{F}}_2}} ~&\!\left\{\!{\varphi _b}\left( {{{\mathbf{F}}_1},{{\mathbf{F}}_2},{t_b^*}} \right) \!-\! \mathop {\max }\limits_k {\varphi _{{e_k}}}\left( {{{\mathbf{F}}_1},{{\mathbf{F}}_2},\!{t_{{e_k}}^*}\!} \right)\!\right\}\!\\ 
	{\rm  s.t.}~~~&\left( {{{\mathbf{F}}_1},{{\mathbf{F}}_2}} \right) \in \mathcal{F}.  \!
	\end{split}
\end{equation*}
\endgroup
Introducing a slack variable $t$, (P1.4) can be equivalently written as
\vspace{-1mm}
\begingroup
\addtolength{\jot}{-1mm}
\begin{equation*}
\small
\vspace{-1mm}
\label{P_1.5}
	\begin{split}
	\left( {{\text{P1.5}}} \right)\!:\mathop {\max }\limits_{{{\mathbf{F}}_1},{{\mathbf{F}}_2},t} ~&\!{\varphi _b}\left( {{{\mathbf{F}}_1},{{\mathbf{F}}_2},{t_b^*}} \right) -t\\ 
	{\rm s.t.}~~~& {\varphi _{{e_k}}}\left( {{{\mathbf{F}}_1},{{\mathbf{F}}_2},\!{t_{{e_k}}^*}} \right)\le t, k=1,...,K,\\
	&\left( {{{\mathbf{F}}_1},{{\mathbf{F}}_2}} \right) \in \mathcal{F}. 
	\end{split}	\vspace{-0mm}	
\end{equation*}
\endgroup
Since (P1.5) is convex, it can be efficiently solved by using a convex optimization solver, e.g. CVX. Note that there is no guarantee that the obtained ${{\bf{F}}_1}$ and ${{\bf{F}}_2}$ are rank-1 matrices as the rank-1 constraints are dropped in (P1.2) by applying SDR. If the obtained ${{\bf{F}}_1}$ and ${{\bf{F}}_2}$ are of rank-1, they can be written as ${{\bf{F}}_1}={{\bf{w}}_1} {{\bf{w}}_1^H}$ and ${{\bf{F}}_2={{\bf{w}}_2} {{\bf{w}}_2^H}}$ by applying eigenvalue decomposition, and then the optimal ${{\bf{f}}_1}$ and ${{\bf{f}}_2}$ are given by ${{\bf{f}}_1}={{\bf{w}}_1}$ and ${{\bf{f}}_2}={{\bf{w}}_2}$, respectively. Otherwise, Gaussian randomization is needed for recovering ${{\bf{f}}_1}$ and ${{\bf{f}}_2}$ approximately, for which the details are omitted \cite{QQ_1}. 

In the above, an approximate solution to (P1.1) is obtained by alternately updating $({{\bf{F}}_1}, {{\bf{F}}_2)}$ and $\left(t_b,t_{e_k}\right)$, which is summarized in Algorithm \ref{Algorithm1}.

\vspace{-4mm}
\subsection{Optimizing ${\bf{v}}$ for Given ${\bf{f}}_1$ and ${\bf{f}}_2$ }
\vspace{-1.5mm}
Next, for any given ${\bf{f}}_1$ and ${\bf{f}}_2$, we denote ${{{\mathbf{\bar h}}}_i} = {{\mathbf{H}}_i}{{\mathbf{f}}_1}$, ${{{\mathbf{\bar H}}}_i} = {{{\mathbf{\bar h}}}_i}{\mathbf{\bar h}}{_i^H}$, ${{{\mathbf{\hat h}}}_i} = {{\mathbf{H}}_i}{{\mathbf{f}}_2}$, and ${{{\mathbf{\hat H}}}_i} = {{{\mathbf{\hat h}}}_i}{\mathbf{\hat h}}_i^H$, $i \in \{b,e_k\}$. As a result, (P0) can be simplified as
\vspace{-1mm}
\begin{equation*}
\small
\vspace{-1mm}
	\label{P_2.1}
	\begin{split}
	\left( {\!{\text{P2.1}}\!} \right) \!:\!
	\mathop {\max }\limits_{\bf{\tilde v}} \
	& \log\!\!\left(\!\!1\!\!+\!\!\!\frac{{{\gamma _0}{{\left| {{{\mathbf{\tilde v}}}^H}{{{\mathbf{\bar h}}}_b} \right|}^2}}}{{{\gamma _0}{{\!\left| {{{\mathbf{\tilde v}}}^H}{{{\mathbf{\hat h}}}_b} \right|\!}^2}\! \!+\!\! 1}}\!\!\!\right)\!\!\!-\!\mathop {\max }\limits_k\, \log\!\!\left(\!\!1\!\!+\!\!\!\frac{{{\gamma _0}{{\left| {{{\mathbf{\tilde v}}}^H}{{{\mathbf{\bar h}}}_{e_k}} \right|}^2}}}{{{\gamma _0}{{\!\left| {{{\mathbf{\tilde v}}}^H}{{{\mathbf{\hat h}}}_{e_k}} \right|\!}^2}\! \!+\!\! 1}}\!\!\!\right)\!\!\notag\\	
	{\rm  s.t.}~~~~&\left| {{v_n}} \right| = 1, n=1,...,N.
	\end{split}
	\vspace{-4mm}
\end{equation*}
Similarly as for (P1.1), by applying Lemma 1 together with SDR, the optimization over ${\bf{\tilde v}}$ for given $({{\bf{f}}_1}, {{\bf{f}}_2)}$ is reduced to\vspace{-1mm}
\begingroup
\addtolength{\jot}{-1mm}
\begin{equation*}
\small
\vspace{-1mm}
\label{P2.2}
	\begin{split}
	\left( {{\text{P2.2}}} \right):\mathop {\max }\limits_{{\mathbf{\tilde V}},{z_b},{z_{e_k}}} ~&\!\left\{{\psi _b}\!\left(\! {{\mathbf{\tilde V}},{z_b}} \right) \!-\! {\mathop {\max }\limits_k {\psi _{{e_k}}}}\!\left( \!{{\mathbf{\tilde V}},{z_{e_k}}} \right)\right\}\! \\ 
	{\rm  s.t.}~~~&{{{\mathbf{\tilde V}}}} \succeq 0, {{{\mathbf{\tilde V}}}_{n,n}}=1, n=1,...,N+1, \\
	&{{\text{z}}_b} > 0,{z_{e_k}} > 0, k=1,...,K, \\ 
	\end{split}	
	\vspace{-2mm}
\end{equation*}
\endgroup
\vspace{-2.5mm}\\
where\vspace{-1mm}
{\small
	\begingroup
	\addtolength{\jot}{-0.1mm}
	\begin{align}
	{\psi _b}\left( {{\mathbf{\tilde V}},{z_b}} \right) = &\ln \left( {{\gamma _0}{\text{Tr}}\left( {\left( {{{{\mathbf{\bar H}}}_b} + {{{\mathbf{\hat H}}}_b}} \right){\mathbf{\tilde V}}} \right) + 1} \right)-  \notag\\
	& {z_b}\left( {{\gamma _0}{\text{Tr}}\left( {{{{\mathbf{\hat H}}}_b}{\mathbf{\tilde V}}} \right) + 1} \right){\text{ + }}\ln {z_b}{\text{ + }}1,	\vspace{-2mm} 
\end{align}
\endgroup
}\vspace{-4mm}\\
and
\vspace{-1.5mm}
{\small
\begingroup
\addtolength{\jot}{-0.5mm}	
\begin{align}
\setlength\belowdisplayskip{-2mm}
	{\psi _{{e_k}}}\left( {{\mathbf{\tilde V}},{z_{{e_k}}}} \right) =& {z_{{e_k}}}\left( {{\gamma _0}{\mathbf{Tr}}\left( {\left( {{{{\mathbf{\bar H}}}_{{e_k}}} + {{{\mathbf{\hat H}}}_{{e_k}}}} \right){\mathbf{\tilde V}}} \right)\! +\! 1} \right) \!-\! \notag\\
	& \ln \left( {{\gamma _0}{\text{Tr}}\left( {{{{\mathbf{\hat H}}}_{{e_k}}}{\mathbf{\tilde V}}} \right) + 1} \right)-\ln {z_{e_k}} - 1. \vspace{-2mm}
\end{align}
\endgroup
\vspace{-4mm}\\
}
It can be verified that (P2.2) is convex w.r.t. either ${\mathbf{\tilde V}}$ or $\left({z_{{b}}},{z_{{e_k}}}\right)$, with the other being fixed. Similarly, it can be approximately solved by alternately optimizing ${{\bf{\tilde V}}}$ and $\left(z_b,z_{e_k}\right)$. For given ${\mathbf{\tilde V}}$, the optimal $\left(z_b,z_{e_k}\right)$ is given by\vspace{-1mm}
{\small
\begingroup
\addtolength{\jot}{-1mm}
\begin{align}
\label{z_b_e}
z_b^*&=\left({{\gamma _0}{\mathbf{Tr}}\!\left(\! {{{{\mathbf{\hat H}}}_b}{\mathbf{\tilde V}}}  \right) \!+\! 1}\right)^{-1},\\
\!z_{e_k}^*&=\left( {{\gamma _0}{\mathbf{Tr}}\left( {\left( {{{{\mathbf{\bar H}}}_{{e_k}}} + {{{\mathbf{\hat H}}}_{{e_k}}}} \right){\mathbf{\tilde V}}} \right)\! +\! 1} \right)^{-1}.\!
\end{align}
\endgroup
\vspace{-4mm}\\
}
While for given $\left(z_b^*,z_{e_k}^*\right)$, the optimal ${{\bf{\tilde V}}}$ is given by\vspace{-1mm}
\begin{equation}
\small
\vspace{0mm}
\label{V}
 {\bf{\tilde V}}^*\!=\! \arg \mathop {\max }\limits_{{{{\mathbf{\tilde V}}}_{n,n}}=1} \left\{{\psi _b}\left( {{\mathbf{\tilde V}},{z_b^*}} \right) \!-\! {\mathop {\max }\limits_k {\psi _{{e_k}}}}\left( {{\mathbf{\tilde V}},{z_{e_k}^*}} \right)\right\},
 \vspace{-0.5mm}
\end{equation}
which can be solved similarly as (P1.5).

After extracting ${{\mathbf{\tilde v}}}$ from ${{\mathbf{\tilde V}}}$ by eigenvalue decomposition with Gaussian randomization, the reflection coefficients are obtained as
\begin{equation}
\setlength\abovedisplayskip{1pt}
\label{v}
	{{{v}}_n} =e^{j\, \angle (\frac{{{{{{\tilde v}}}_n}}}{{{{{{\tilde v}}}_{{N + 1}}}}})}, n = 1,...,N,
	\vspace{-2mm}
\end{equation}
where $\angle(x)$ denotes the phase of $x$ and the constraints $\left| {{v_n}} \right| = 1$, $\forall n$, are satisfied.

\vspace{-3mm}
\subsection{Overall Algorithm}
\vspace{-0.5mm}
To summarize, the overall iterative algorithm to solve (P0) is given in Algorithm \ref{Algorithm2}, where $\epsilon$ denotes a small threshold and $L$ is the maximum number of iterations. 

\setlength{\textfloatsep}{4mm} 
\begin{algorithm}[t]
	\caption{Alternating optimization for solving (P0)}
	\label{Algorithm2}
	\LinesNumbered 
	\DontPrintSemicolon	
	\SetAlgoHangIndent{0.1em}
	\KwIn{$P_{\rm max}$, $\gamma_0$, ${\bf{H}}_{b}$, ${\bf{H}}_{{e_k}}$, $\epsilon$, $L$.}
	\KwOut{$\bf{f_1}$, $\bf{f_2}$, ${\bf{v}}$. }
	Initialize the reflection coefficients vector as ${\mathbf{v}}^{(0)}$. \\
	Set $l=1$, ${{{\mathbf{\tilde v}}}^{(0)}} = \!\left[\! {\begin{array}{*{20}{c}}
							{{\mathbf{v}}^{(0)}} \\ 
							1 
							\end{array}} \!\right]\!$.\\ 
	\Repeat{\rm the fractional decrease of the objective value in (P0) is below $\epsilon$ or $l=L$ .}
	{
		Solve (P1.1) for given ${\bf{\tilde v}}^{(l-1)}$ by applying Algorithm \ref{Algorithm1}, and denote the solution as ${\bf{f}}_1^{(l)}$ and ${\bf{f}}_2^{(l)}$.
		\\
		Solve (P2.1) for given ${\bf{f}}_1^{(l)}$ and ${\bf{f}}_2^{(l)}$, and denote the solution as ${\bf{\tilde v}}^{(l)}$.\\	
		Update $l=l+1$.
	}	        
	Recover ${\bf{v}}$ from ${\bf{\tilde v}}$ according to \eqref{v}.\vspace{-0mm}
\end{algorithm}

\vspace{-0mm}
\section{Simulation Results}
The simulation setups are shown in Fig. 2. It is assumed that Alice, Rose (the central point) and Bob are located at (5, 0, 20), (0, 100, 2), and (3, 100, 0) in meter (m), respectively. To study the effect of jamming, we consider two different setups in Fig. 2 where the $K$ Eves lie uniformly along the line from (2, 95, 0) to (2, 105, 0) in Setup (a) and from (2, $-$105, 0) to (2, $-$95, 0) in Setup (b), thus corresponding to the cases with local Eves and remote Eves near/from the IRS, respectively. The simulation parameters are set as shown in Table \ref{parameters}.

The channel from Alice to Bob is generated by ${{\mathbf{h}}^H_{ab}} = \sqrt {{L_0}d_{ab}^{ - {c_{ab}}}} {{\bf{g}}_{ab}}$, where $d_{ab}$ denotes the distance from Alice to Bob and ${{\bf{g}}_{ab}}$ is the small-scale fading component assumed to be Rician fading and given by  \vspace{0mm}
\begin{equation}
	{{\bf{g}}_{ab}} = \sqrt {{{{\beta _{ab}}}}/({{1 + {\beta _{ab}}}})} {\bf{g}}_{ab}^{\rm LoS} + \sqrt {{1}/({{1 + {\beta _{ab}}}})}{\bf{g}}_{ab}^{\rm NLoS} ,\vspace{-1mm}
\end{equation}
where ${\bf{g}}_{ab}^{\rm LoS}$ and ${\bf{g}}_{ab}^{\rm NLoS}$ represent the deterministic line-of-sight (LoS) and Rayleigh fading/non-LoS (NLoS) components, respectively. The same channel model is adopted for ${{\mathbf{h}}^H_{a{e_k}}}$, ${{\mathbf{h}}^H_{r{b}}}$, ${{\mathbf{h}}^H_{r{e_k}}}$ and ${\bf{H}}_{ar}$. We assume that the channels from Alice to Bob, Rose, and Eve $k$ have no LoS component and experience Rayleigh fading. Considering that Rose is deployed vertically higher than Bob and Eves, a less scattering environment is expected and thus we set $c_{ar}<c_{ai}, i \in \{{b,e_k}$\}. In Setup (a), we assume that the channels from Rose to Bob and Eve $k$ are LoS, while in Setup (b), we assume that the channel from Rose to Eve $k$ experiences Rayleigh fading. 
\begin{figure}[!t]
	\label{Sim_model}
	\centering{\includegraphics[width=3.7in]{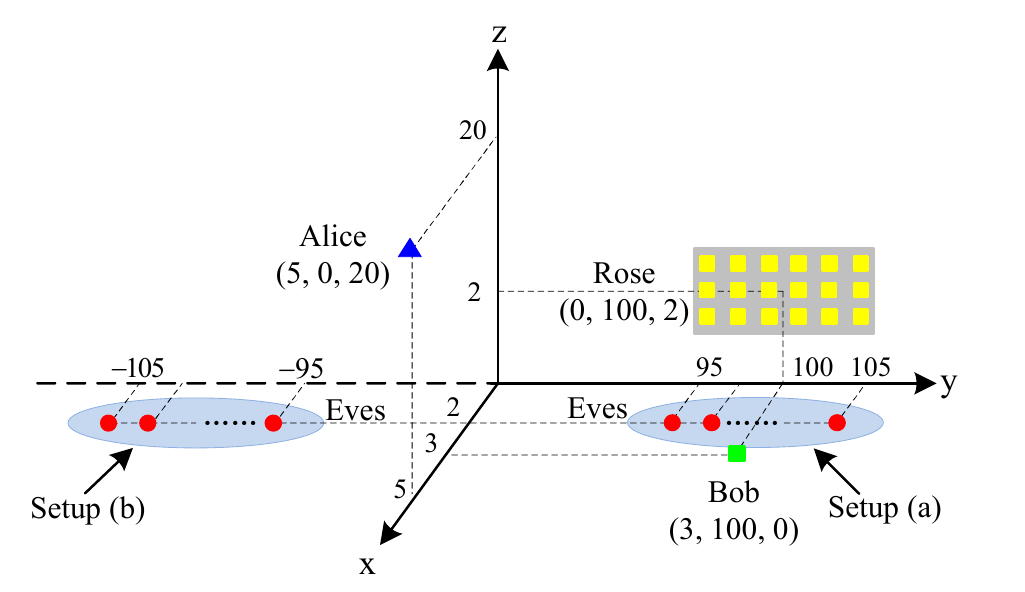}}
	\caption{Simulation setups. }
	\vspace{-0mm}
\end{figure}
\begin{table}[!t]	
	\small
	\renewcommand{\arraystretch}{1.3}
	\caption{Simulation Parameters}
	\label{parameters}
	\centering
	\begin{tabular}{|p{2.4cm}|p{5.3cm}|}
		\hline 
		\textbf{Parameter} & \textbf{Value}\\
		\hline
		Carrier frequency  & 750 MHz.\\
		\hline
		IRS configuration  & Uniform rectangular array (URA) with 5 rows and ${N}/{5}$ columns, ${3\lambda}/{8}$ spacing.\\
		\hline
		Path loss at 1m   & $L_0=-30$ dB. \\
		\hline
		Path loss exponent & $\!c_{ab}\!=\!c_{a{e_k}}\!\!=\!5$, $\!c_{ar}\!\!=\!3.5$, $\!c_{rb}\!\!=\!2$, $\!c_{r{e_k}}\!\!=\!2$ and 5 for Setup (a) and (b), respectively.  \\
		\hline
		Racian factor      & $\!\beta_{ab}\!\!=\!\!\beta_{a{e_k}}\!\!=\!\!\beta_{ar}\!\!=\!0$, $\!\beta_{rb}\!\!=\!\infty$, $\!\beta_{r{e_k}}\!\!=\!\infty$ and 0 for Setup (a) and (b), respectively.\\
		\hline
		Other parameters   & $\sigma_0^2=-105$ dBm, $\epsilon=10^{-3}$, $L=40$.\\
		\hline	
	\end{tabular}
\end{table}

In addition to the proposed design for the case with IRS and AN (AN, IRS), other cases including with AN but without IRS (AN, No-IRS) \cite{LQ_2013}, with IRS but without AN (No-AN, IRS), and without both IRS and AN (No-AN, No-IRS) are also adopted for performance comparison. Note that by setting ${\bf{f}}_2={\bf{0}}$ (i.e., the case of No-AN, IRS) and $K=1$, the setup is the same as that considered in \cite{Guangchi}. 

The achievable secrecy rate versus the transmit power of Alice is plotted in Fig. 3. It can be observed that as the transmit power increases, the AN-aided designs outperform their counterparts without AN, for both the cases with and without the IRS in both Setups (a) and (b). Note that the achievable secrecy rates for both Setups (a) and (b) are identical for the cases without IRS due to the symmetry of Eves' locations at the two sides of Alice. In fact, as $P_{\rm max}$ goes to $\infty$, $(1+\gamma_b)/(1+\mathop {\max }\limits_k\gamma_{e_k})$ converges to a constant, which implies that  increasing transmit power alone is inefficient for improving the secrecy rate and incorporating AN is beneficial.

Fig. 4 shows the secrecy rate gains achieved by using AN with increasing the number of Eves, $K$. Note that when $K = 1$, the secrecy rates with and without AN are almost the same, regardless of whether IRS is used or not. This is expected because the number of transmit antennas is much larger than that of Eves and thus transmit beamforming has sufficient spatial DoF to suppress the signal in the Eves' direction, rendering the use of AN unnecessary. However, as the number of Eves increases, transmit beamforming lacks sufficient DoF for signal nulling and thus it becomes more beneficial to allocate part of transmit power to send jamming signal for degrading the reception of Eves. Interestingly, in Setup (a), it is observed that the case of (AN, No-IRS) even outperforms that of (No-AN, IRS) when $K \ge 6$. This implies that, in this more challenging setup with both Bob and Eves near IRS, AN is particularly useful, as the additional DoF provided by IRS may be insufficient to prevent the information leakage to Eves due to their proximity to the IRS as Bob. 

\begin{figure}[!t]
	\label{Cs_Pmax_E1E3}
	\centering{\includegraphics[width=2.6in]{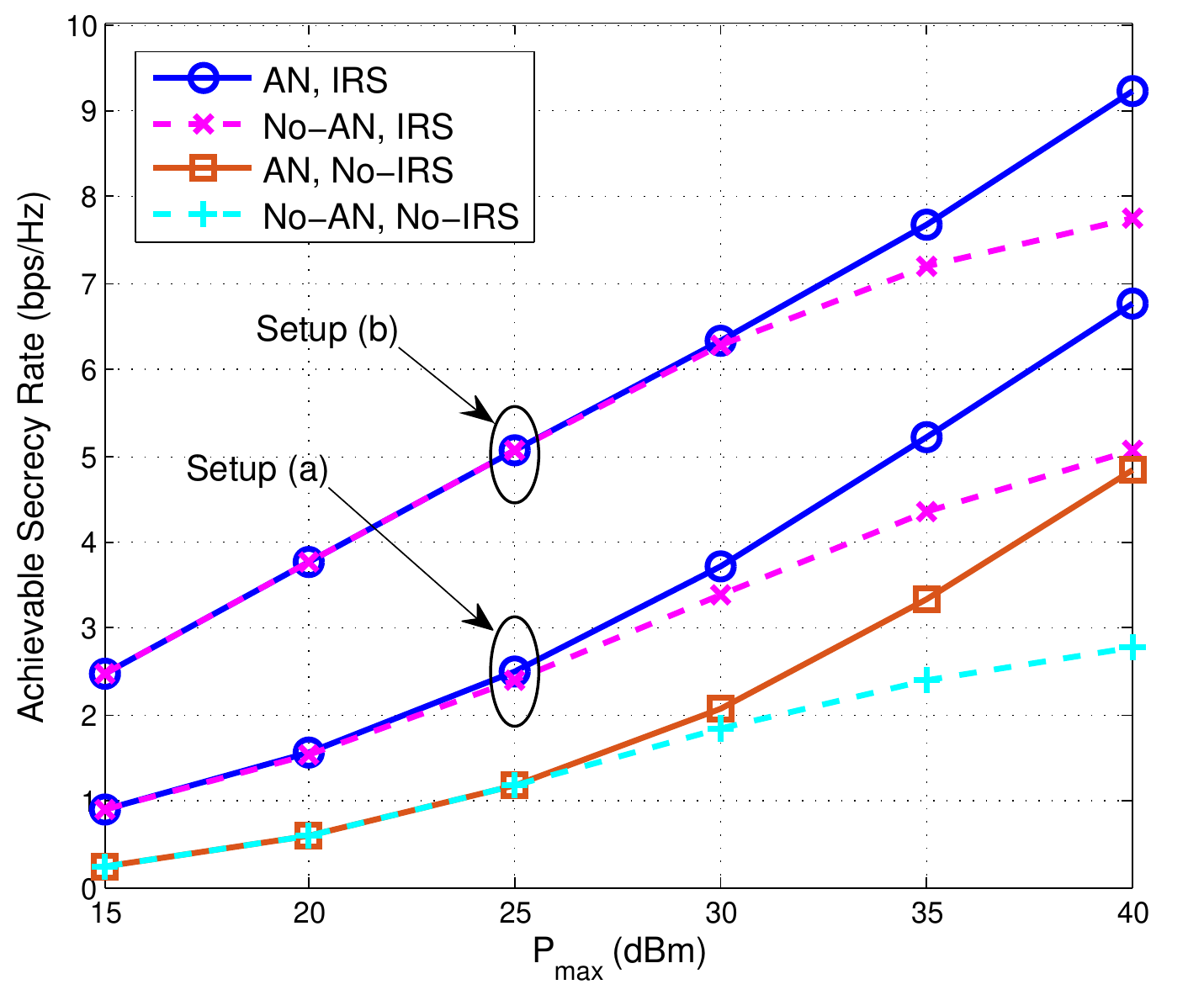}}
	\caption{Achievable secrecy rate versus the maximum transmit power, $P_{\rm max}$, with $(M, N, K)=$ (4, 20, 5).}
	\vspace{0.3mm}
\end{figure}

\begin{figure}[!t]
	\label{Cs_K_E1E3}
	\centering{\includegraphics[width=2.63in]{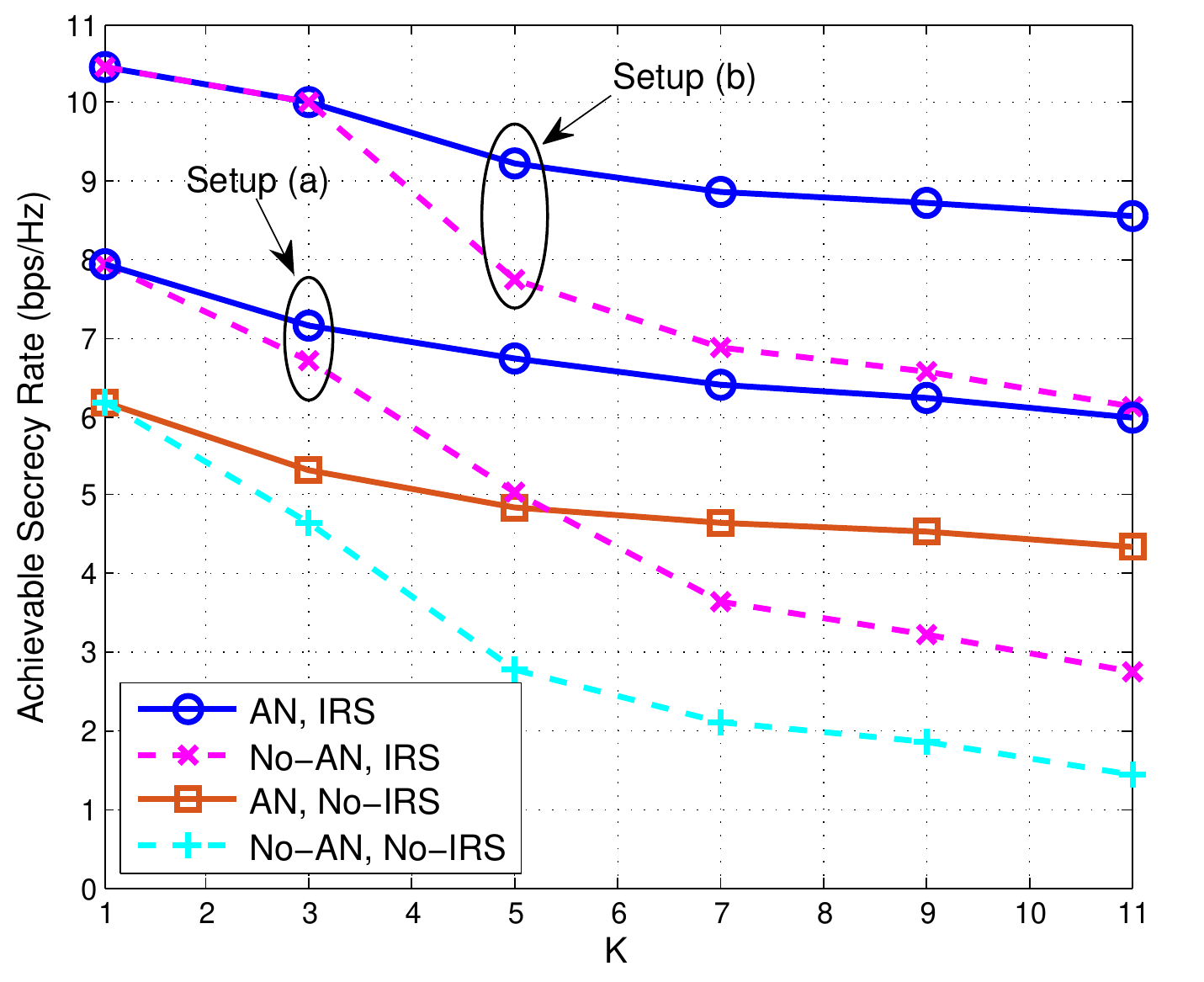}}
	\caption{Achievable secrecy rate versus the number of Eves, $K$, with $(M, N, P_{\rm max})=$ (4, 20, 40~\rm {dBm}).}
\end{figure}

Fig. 5 depicts the achievable secrecy rate versus the number of reflecting elements of the IRS, $N$. It is observed that even with IRS, the AN-aided design requires less reflecting elements to achieve the same secrecy rate as compared to the No-AN design. It is also observed that the performance gain by using AN decreases with increasing $N$ in Setup (a), while it remains almost unchanged in Setup (b). This is expected since in Setup (a), more DoF become available for the passive beamforming of the IRS with larger $N$ to degrade the reception at the Eves, which thus renders the use of AN less effective. However, when the Eves are far away from the IRS in Setup (b), the reflect beamforming of the IRS is fully exploited to enhance the desired signal at the Bob's receiver, but without the need of nulling/canceling the signals at the Eves that are out of its coverage. As a result, the performance gain due to AN is roughly constant regardless of $N$.  

Finally, it is observed from Figs. 3-5 that Setup (b) always achieves higher secrecy rate than Setup (a) for the case with IRS, regardless of whether AN is used or not. The reason is that in Setup (a), the Eves are in the same local region as Bob covered by Rose (IRS), and as a result it becomes more challenging to degrade the reception of the Eves, for the design of both transmit beamforming with/without AN and reflect beamforming of the IRS. 
\setlength{\textfloatsep}{4mm} 
\begin{figure}[!t]
	\label{CsVsN_E1E3}
	\centering{\includegraphics[width=2.63in]{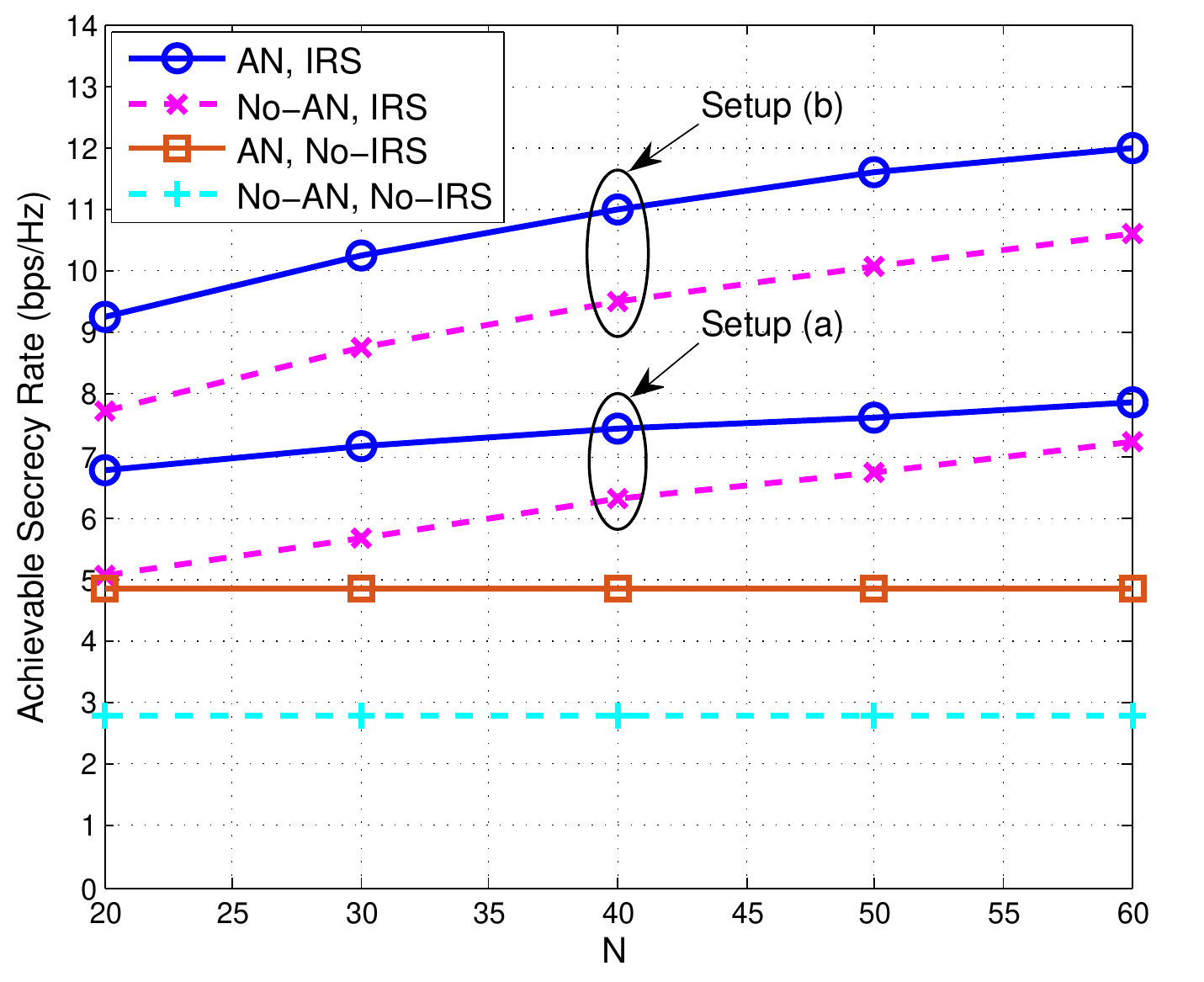}}
	\caption{Achievable secrecy rate versus the number of reflecting elements of the IRS, $N$, with $(M, K, P_{\rm max})=$ (4, 5, 40 dBm).}
\end{figure}

\section{Conclusion}
In this letter, we investigated whether AN is helpful to enhance the physical layer security in the new IRS-assisted communication system. To answer this question, we formulated a secrecy rate maximization problem for the joint design of transmit/reflect beamforming with AN. An alternating optimization based algorithm was developed to solve this problem efficiently. By simulation results, we verified the necessity of using AN even with an IRS deployed and identified the practical scenarios when the use of AN is most beneficial. It was shown that transmit and reflect beamforming alone in general cannot deal with increasing number of eavesdroppers effectively due to the lack of sufficient spatial DoF, while AN can be an effective means to help improve the secrecy rate even in such challenging case.

\ifCLASSOPTIONcaptionsoff
\newpage
\fi

\bibliographystyle{IEEEtran}
\bibliography{IEEEabrv,reference}

\end{document}